\documentclass[onecolumn,showpacs,preprintnumbers,amsmath,aps]{revtex4}
\usepackage{graphicx}
\usepackage{dcolumn}
\usepackage{bm}
\usepackage{color}
\begin{document}
%
\def\eq#1{(\ref{#1})}
\def\fig#1{Fig.\hspace{1mm}\ref{#1}}
\def\tab#1{\hspace{1mm}\ref{#1}}
\title{
---------------------------------------------------------------------------------------------------------------\\
The characterization of high-pressure superconducting state in ${\rm Si_{2}H_{6}}$ compound: the strong-coupling description}
\author{R. Szcz{\c{e}}{\`s}niak, A.P. Durajski}
\affiliation{Institute of Physics, Cz{\c{e}}stochowa University of Technology, Al. Armii Krajowej 19, 42-200 Cz{\c{e}}stochowa, Poland}
\date{\today} 
\begin{abstract}
The thermodynamic parameters of the superconducting state, that induces in ${\rm Si_{2}H_{6}}$ under the pressure at $275$ GPa, have been calculated. In the framework of the Eliashberg formalism, it has been shown that the critical temperature can attain extremely high values: $T_{C}\left(\mu^{\star}\right)\in\left<173.36, 99.99\right>$ K, where the Coulomb pseudopotential $\left(\mu^{\star}\right)$ belongs to the range from $0.1$ to $0.3$. The ratio of the energy gap to the critical temperature ($R_{\Delta}$) significantly exceeds the value predicted by the BCS theory: $R_{\Delta}\left(\mu^{\star}\right)\in\left<4.40, 4.04\right>$. Additionally, it has been stated that in the whole range of the superconducting state's existence, the electron effective mass ($m^{\star}_{e}$) is large; $\left[m^{\star}_{e}\right]^{\rm max}_{T=T_{C}}=2.397m_{e}$, where the symbol $m_{e}$ denotes the electron band mass. 
\end{abstract}
\maketitle

The induction of the superconducting state with the high value of the critical temperature ($T_{C}$) is one of the fundamental problems of the modern solid state physics. 

The conducted investigations allowed to determine that the electron-phonon interaction in the two-band system could lead to the formation of the superconducting phase with the relatively high critical temperature; the maximum of $T_{C}$ has been discovered for $\rm MgB_{2}$ (under normal conditions), where the critical temperature is equal to $39.4$ K \cite{Nagamatsu}. It should be noticed that the remaining thermodynamic parameters of the superconducting state differ significantly from the predictions of the BCS theory \cite{Golubov}, \cite{Szczesniak1}, \cite{BCS1}, \cite{BCS2}. 

As for now, the highest values of the critical temperature can be observed in the cuprates discovered in 1986 by Bednorz and Muller \cite{Bednorz}. In particular, for the compound ${\rm HgBa_{2}Ca_{2}Cu_{3}O_{8+\delta}}$ under the normal conditions, $T_{C}=135$ K has been stated \cite{Chu}. Under the influence of the pressure $p\simeq 31$ GPa, the critical temperature in ${\rm HgBa_{2}Ca_{2}Cu_{3}O_{8+\delta}}$ increases up to the value of about $164$ K \cite{GaoHTSC}. Due to the lack of the generally acceptable theory of the superconducting state in the cuprates, it is difficult to say how high value may take $T_{C}$. However, some clues can be provided by the recently achieved results which are presented in the paper \cite{Szczesniak2}.

Another direction of the studies is being connected with the possibility of the induction of the high-temperature superconducting state in the classical systems that stay under the influence of the high pressure. The strong argument for this type of research are the experimental results obtained for lithium and calcium. In the case of lithium, it has been found $\left[T_{C}\right]_{\rm max}=14$ K for $p=30.2$ GPa, whereas for calcium $\left[T_{C}\right]_{\rm max}$ equals $25$ K  ($p=161$ GPa) \cite{Deemyad}, \cite{Yabuuchi}. However, Sakata {\it et al.} have suggested that the value of the critical temperature for calcium can reach $29$ K ($p=216$ GPa) \cite{Sakata} - but this result has been challenged by Andersson \cite{Andersson}. A full review of an anomalous thermodynamic properties of the superconducting state in lithium, calcium and ${\rm CaLi_{2}}$ can be found by the readers in papers: \cite{Szczesniak3}, \cite{Szczesniak4}, \cite{Szczesniak5}, \cite{Szczesniak6}, \cite{Szczesniak7}, \cite{Szczesniak8}.     

From the theoretical point of view, the highest value of the critical temperature should characterize the superconducting state in the metallic hydrogen \cite{Ashcroft}. The numerical calculations performed in the pressure range from $\sim 400$ GPa to $3.5$ TPa revealed that $T_{C}$ is extremely high \cite{Cudazzo}, \cite{Szczesniak9}, \cite{Szczesniak10}, \cite{Maksimov}, \cite{McMahon}. In particular, for $p=2$ TPa the critical temperature changes from $413$ K to $631$ K (depending on the assumed Coulomb pseudopotential) \cite{Szczesniak11}. 

The above theoretical results, although very interesting, are now experimentally unverifiable because of the high pressure metallization 
($p_{\rm m}\sim 400$ GPa) \cite{Stadele}. For this reason, the way for the reduction of $p_{{\rm m}}$ is being sought out. A good suggestion seems to be the use of the chemical pre-compression \cite{Feng}. The recently conducted studies allowed to figure out that the metallization pressure for ${\rm SiH_{4}}$ is about $55$ GPa \cite{Chen}, \cite{Eremets}. What is more important - the compound ${\rm SiH_{4}}$ goes into the superconducting state with the critical temperature of $17$ K for the pressure $96$ GPa and $120$ GPa \cite{Eremets}. 
The theoretical calculations suggest even higher values of the critical temperature for the system ${\rm Si_{2}H_{6}}$ crystallizing in the structure $Pm-3m$ at $275$ GPa \cite{Jin} and for the compound ${\rm SiH_4(H_2)_2}$ - the structure $Ccca$ at $250$ GPa \cite{Li}. 

In the paper, the thermodynamic parameters of the superconducting state in ${\rm Si_{2}H_{6}}$ ($p=275$ GPa) have been determined. Due to the high value of the electron-phonon coupling constant ($\lambda=1.4$), the calculations have been carried out in the framework of the strong-coupling formalism (the Eliashberg approach) \cite{Eliashberg}, \cite{Allen}, \cite{Carbotte01}, \cite{Carbotte02}.

\vspace*{0.5cm}

The Eliashberg equations in the mixed representation have the following form \cite{Marsiglio}:

%
\begin{widetext}
\begin{eqnarray}
\label{r1}
\phi\left(\omega+i\delta\right)&=&
                                  \frac{\pi}{\beta}\sum_{m=-M}^{M}
                                  \left[\lambda\left(\omega-i\omega_{m}\right)-\mu^{\star}\theta\left(\omega_{c}-|\omega_{m}|\right)\right]
                                  \frac{\phi_{m}}
                                  {\sqrt{\omega_m^2Z^{2}_{m}+\phi^{2}_{m}}}\\ \nonumber
                              &+& i\pi\int_{0}^{+\infty}d\omega^{'}\alpha^{2}F\left(\omega^{'}\right)
                                  \left[\left[N\left(\omega^{'}\right)+f\left(\omega^{'}-\omega\right)\right]
                                  \frac{\phi\left(\omega-\omega^{'}+i\delta\right)}
                                  {\sqrt{\left(\omega-\omega^{'}\right)^{2}Z^{2}\left(\omega-\omega^{'}+i\delta\right)
                                  -\phi^{2}\left(\omega-\omega^{'}+i\delta\right)}}\right]\\ \nonumber
                              &+& i\pi\int_{0}^{+\infty}d\omega^{'}\alpha^{2}F\left(\omega^{'}\right)
                                  \left[\left[N\left(\omega^{'}\right)+f\left(\omega^{'}+\omega\right)\right]
                                  \frac{\phi\left(\omega+\omega^{'}+i\delta\right)}
                                  {\sqrt{\left(\omega+\omega^{'}\right)^{2}Z^{2}\left(\omega+\omega^{'}+i\delta\right)
                                  -\phi^{2}\left(\omega+\omega^{'}+i\delta\right)}}\right],
\end{eqnarray}
and
\begin{eqnarray}
\label{r2}
Z\left(\omega+i\delta\right)&=&
                                  1+\frac{i}{\omega}\frac{\pi}{\beta}\sum_{m=-M}^{M}
                                  \lambda\left(\omega-i\omega_{m}\right)
                                  \frac{\omega_{m}Z_{m}}
                                  {\sqrt{\omega_m^2Z^{2}_{m}+\phi^{2}_{m}}}\\ \nonumber
                              &+&\frac{i\pi}{\omega}\int_{0}^{+\infty}d\omega^{'}\alpha^{2}F\left(\omega^{'}\right)
                                  \left[\left[N\left(\omega^{'}\right)+f\left(\omega^{'}-\omega\right)\right]
                                  \frac{\left(\omega-\omega^{'}\right)Z\left(\omega-\omega^{'}+i\delta\right)}
                                  {\sqrt{\left(\omega-\omega^{'}\right)^{2}Z^{2}\left(\omega-\omega^{'}+i\delta\right)
                                  -\phi^{2}\left(\omega-\omega^{'}+i\delta\right)}}\right]\\ \nonumber
                              &+&\frac{i\pi}{\omega}\int_{0}^{+\infty}d\omega^{'}\alpha^{2}F\left(\omega^{'}\right)
                                  \left[\left[N\left(\omega^{'}\right)+f\left(\omega^{'}+\omega\right)\right]
                                  \frac{\left(\omega+\omega^{'}\right)Z\left(\omega+\omega^{'}+i\delta\right)}
                                  {\sqrt{\left(\omega+\omega^{'}\right)^{2}Z^{2}\left(\omega+\omega^{'}+i\delta\right)
                                  -\phi^{2}\left(\omega+\omega^{'}+i\delta\right)}}\right]. 
\end{eqnarray}
\end{widetext}
%

The symbols $\phi\left(\omega\right)$, $Z\left(\omega\right)$ and ($\phi_{m}\equiv\phi\left(i\omega_{m}\right)$, $Z_{m}\equiv Z\left(i\omega_{m}\right)$) denote the order parameter function and the wave function renormalization factor on the real (imaginary) axis, respectively; $\omega_{m}$ represents the $m$-th Matsubara frequency: $\omega_{m}\equiv \left(\pi /\beta\right)\left(2m-1\right)$, where $\beta\equiv\left(k_{B}T\right)^{-1}$ ($k_{B}$ is the Boltzmann constant). The order parameter is defined with an expression: $\Delta\equiv \phi/Z$.

The functions $\phi_{m}$ and $Z_{m}$ should be calculated by solving the Eliashberg equations on the imaginary axis:

\begin{equation}
\label{r3}
\phi_{m}=\frac{\pi}{\beta}\sum_{n=-M}^{M}
\frac{\lambda\left(i\omega_{m}-i\omega_{n}\right)-\mu^{\star}\theta\left(\omega_{c}-|\omega_{n}|\right)}
{\sqrt{\omega_n^2Z^{2}_{n}+\phi^{2}_{n}}}\phi_{n},
\end{equation}
\begin{equation}
\label{r4}
Z_{m}=1+\frac{1}{\omega_{m}}\frac{\pi}{\beta}\sum_{n=-M}^{M}
\frac{\lambda\left(i\omega_{m}-i\omega_{n}\right)}{\sqrt{\omega_n^2Z^{2}_{n}+\phi^{2}_{n}}}\omega_{n}Z_{n}.
\end{equation}

The pairing kernel for the electron-phonon interaction is given by:
$\lambda\left(z\right)\equiv 2\int_0^{\Omega_{\rm{max}}}d\Omega\frac{\Omega}{\Omega ^2-z^{2}}\alpha^{2}F\left(\Omega\right)$,
where $\alpha^{2}F\left(\Omega\right)$ means the Eliashberg function. For ${\rm Si_{2}H_{6}}$, the Eliashberg function has been calculated in \cite{Jin}; the maximum phonon frequency ($\Omega_{\rm{max}}$) is equal to $284.41$ meV.

The Coulomb interaction, occurring between the electrons, is parametrized using the Coulomb pseudopotential $\mu^{\star}$. The symbol $\theta$ denotes the Heaviside function; $\omega_{c}$ is the cut-off frequency ($\omega_{c}=3\Omega_{\rm{max}}$).

The symbol $N\left(\omega\right)$ and $f\left(\omega\right)$ represents the Bose-Einstein and Fermi-Dirac function, respectively. The Eliashberg equations have been solved with the help of the numerical methods used in the papers: \cite{Szczesniak12}, \cite{Szczesniak13}, \cite{Szczesniak14}, \cite{Szczesniak15}. The convergence of the solutions has been obtained for $T\geq T_{0}=23.21$ K ($M=1100$). 

\vspace*{0.5cm}

%
\begin{figure*}[th]
\includegraphics[scale=0.60]{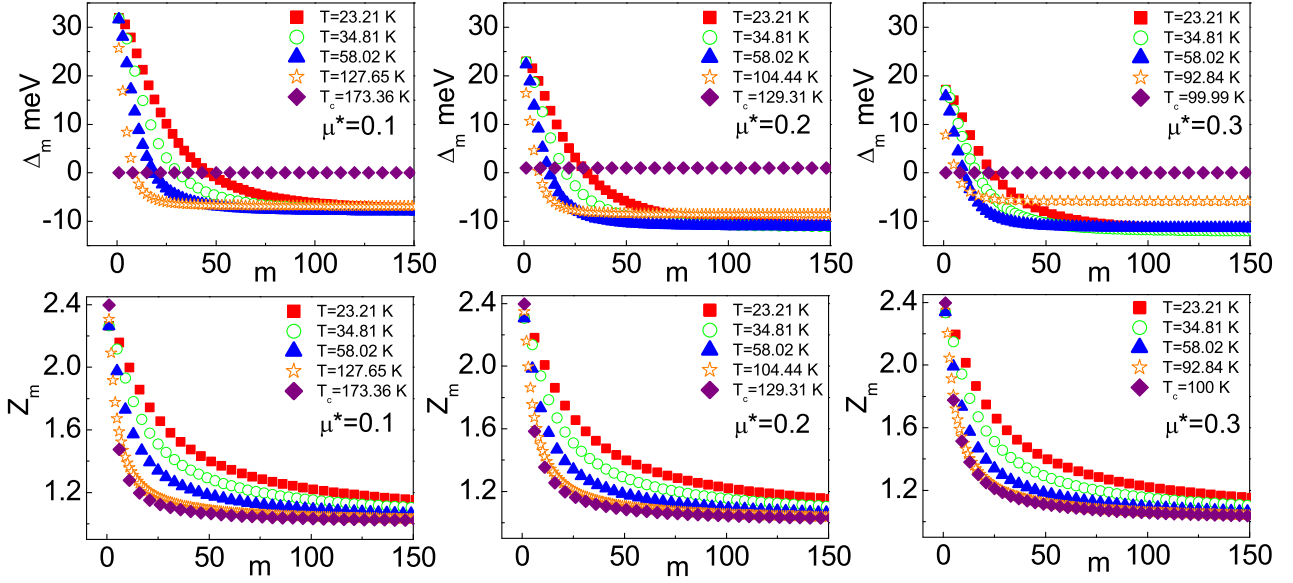}
\caption{
The solutions of the Eliashberg equations on the imaginary axis for the selected values of the temperature and the Coulomb pseudopotential. The first $150$ values of $\Delta_{m}$ and $Z_{m}$ have been shown.}
\label{f1}
\end{figure*}
%

%
\begin{figure}[th]
\includegraphics[scale=0.30]{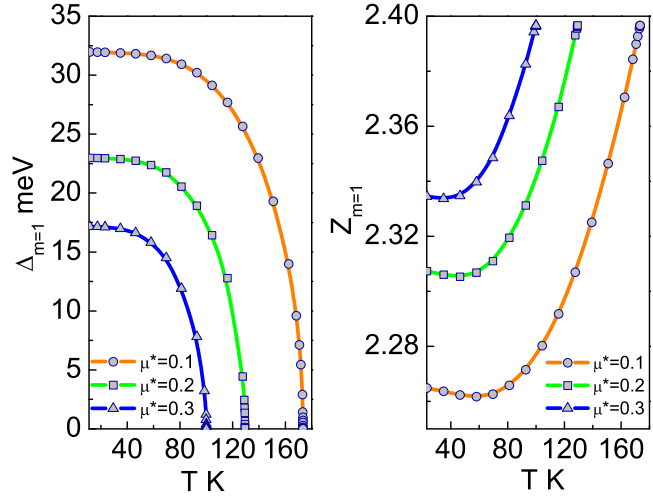}
\caption{
The dependence of $\Delta_{m=1}$ and $Z_{m=1}$ on the temperature for the selected values of the Coulomb pseudopotential.}
\label{f2}
\end{figure}
%

The solutions of the Eliashberg equations have been analyzed for $\mu^{\star}\in\left<0.1,0.3\right>$. The results obtained for the imaginary axis have been presented in \fig{f1}. In particular, in the first row there is plotted a dependence of the order parameter on the temperature and the Coulomb pseudopotential. The second row presents the results for the wave function renormalization factor.

On the basis of \fig{f1}, it is easy to notice that the values of the function $\Delta_{m}$ strongly decrease together with the increase of $m$. The increase of the temperature and the Coulomb pseudopotential causes also a decrease in the value of the parameter $\Delta_{m}$.     

In the case of the wave function renormalization factor, the growth of the number $m$ results in the decrease in the value of $Z_{m}$. However, the wave function renormalization factor very weakly depends on the temperature and the Coulomb pseudopotential.

The full dependence of the order parameter and the wave function renormalization factor on the temperature for the selected values of the Coulomb pseudopotential can be most conveniently traced after plotting the shape of the function $\Delta_{m=1}$ and $Z_{m=1}$ (see \fig{f2}). In particular,  the maximum value of the order parameter can be parametrized with the help of the following expression: $\Delta_{m=1}\left(T,\mu^{\star}\right)=\Delta_{m=1}\left(T_{0},\mu^{\star}\right)\sqrt{1-\left(\frac{T}{T_{C}}\right)^{\beta}}$, where $\Delta_{m=1}\left(T_{0},\mu^{\star}\right)=157.54\left(\mu^{\star}\right)^{2}-137.11\mu^{\star}+44.11$ meV and $\beta=3.4$.

\begin{figure}%
\includegraphics[scale=0.30]{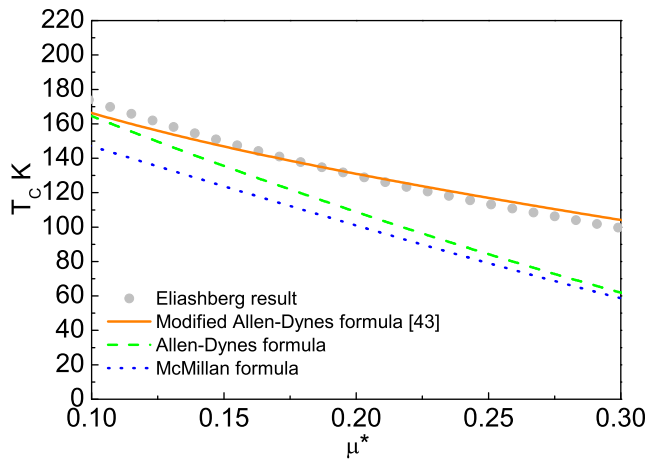}
\caption{The dependence of the critical temperature on the values of the Coulomb pseudopotential, determined
with an use of the selected methods.}
\label{f3}
\end{figure}

In the next step, the dependence of $T_{C}$ on $\mu^{\star}$ has been precisely determined. On the basis of the results presented in \fig{f3}, it is easy to notice that the critical temperature is very high in the whole range of the considered values of the Coulomb pseudopotential; $T_{C}\in\left<173.36, 99.99\right>$ K. 
Let us note that the calculated critical temperatures ($T_{C}\left(\mu^{\star}=0.1\right)=173.36$ K and $T_{C}\left(\mu^{\star}=0.13\right)=158.57$ K) significantly exceed the values determined in the paper \cite{Jin}: $T_{C}\left(\mu^{\star}=0.1\right)=153.44$ K and $T_{C}\left(\mu^{\star}=0.13\right)=138.86$ K. The obtained result is related to the fact that $T_{C}$ in \cite{Jin} has been estimated on the basis of the classical Allen-Dynes formula, which significantly lowers its value.

Generalizing the achieved result, \fig{f3} presents the complete dependence of the critical temperature on the Coulomb pseudopotential calculated on the basis of the classical expressions of Allen-Dynes and McMillan \cite{AllenDynes}, \cite{McMillan}. It can be clearly seen that with the increase of $\mu^{\star}$ the critical temperature, based on classical formulas, more visibly differs from the strict result based on the Eliashberg equations.

Let us notice that the function $T_{C}\left(\mu^{\star}\right)$ can be determined with a good approximation using the analytical formula. However, it is necessarry to use the formula originally derived for ${\rm SiH_4(H_2)_2}$ \cite{wzor1}.

%
%

\vspace*{0.5cm}

The Eliashberg equations in the mixed representation have been solved for the identical range of the temperatures and the Coulomb pseudopotential as the Eliashberg equations on the imaginary axis.

\begin{figure*}[th]
\includegraphics[scale=0.55]{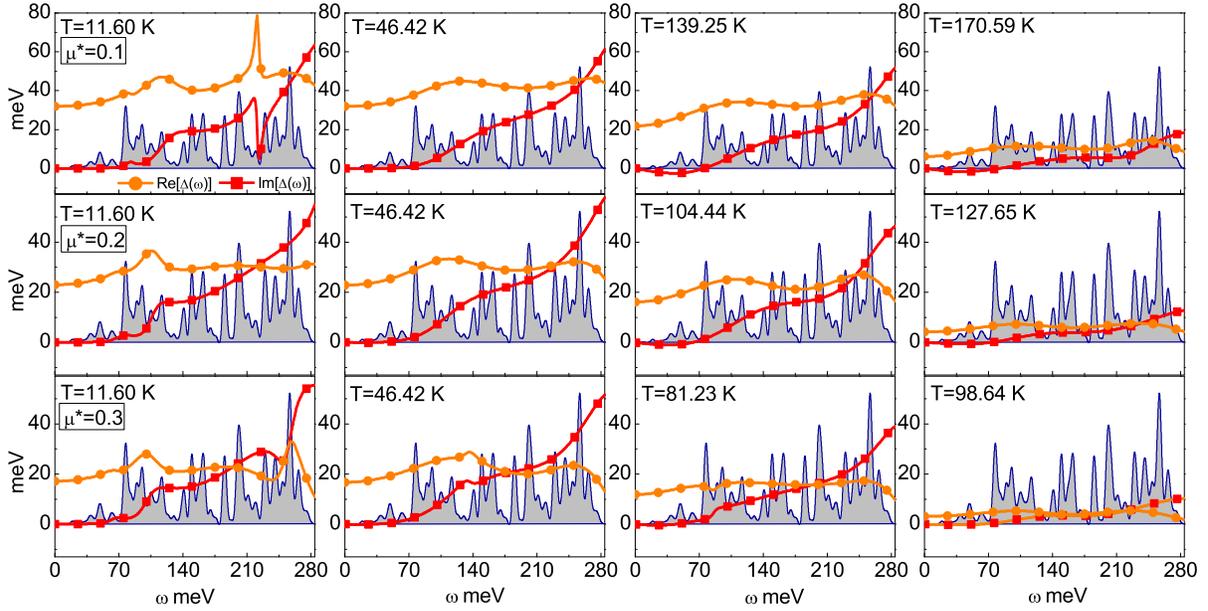}
\caption{ The real and imaginary part of the order parameter on the real axis for the selected values of the
temperature and the Coulomb pseudopotential. Additionally, the rescaled Eliashberg function has been plotted.} 
\label{f4}
\end{figure*}

\fig{f4} presents the form of the order parameter on the real axis in the frequency range from $0$ to $\Omega_{\rm max}$. Additionally, the rescaled Eliashberg function ($30\alpha^{2}F\left(\Omega\right)$) has been also plotted. It can be easily seen that for low temperatures, the course of the order parameter is far more complex than for the higher temperatures. The difference stems from the fact that in the range of the low temperatures the form of the function $\Delta\left(\omega\right)$ is strongly correlated with the complicated shape of the Eliashberg function \cite{Varelogiannis}. 

\begin{figure*}[th]
\includegraphics[scale=0.60]{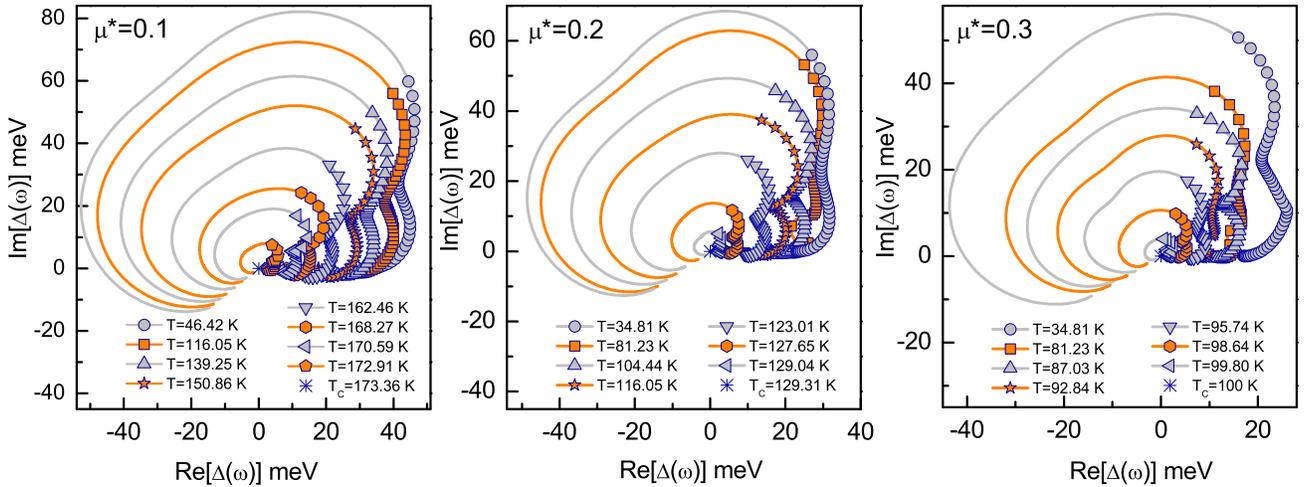}
\caption{
The order parameter on the complex plane for the selected values of the temperature and the Coulomb
pseudopotential. The lines with symbols have been achieved for $\omega\in\left<0,\Omega_{{\rm max}}\right>$; the lines without symbols
correspond to the frequencies' range from $\Omega_{{\rm max}}$ to $\omega_{c}$}. 
\label{f5}
\end{figure*}

It is also convenient to plot the order parameter on the complex plane. In particular, \fig{f5} presents the values of $\Delta\left(\omega\right)$ in the dependence on the temperature and the Coulomb pseudopotential. The wide range of the frequencies has been selected: $\omega\in\left<0,\omega_{c}\right>$. It can be noticed that the values of the order parameter lay down on the characteristic spirals with the radius decreasing together with the increase of $T$ and $\mu^{\star}$. Note that similar spirals for the order parameter have been observed for Pb, Hg and Sn in the paper \cite{spirale}.

The courses plotted in \fig{f5} allow to determine the frequency range for which the effective potential of the electron-electron interaction is pairing. From the mathematical point of view, the values of the above frequencies are calculated on the basis of the following condition: Re$\left[\Delta\left(\omega\right)\right]>0$ \cite{Varelogiannis}. For $\mu^{\star}=0.1$, the range of the frequency corresponding to the pairing potential extends from $0$ to $\omega_{p}\simeq 1.2\Omega_{{\rm max}}$. With the increase of the Coulomb pseudopotential $\omega_{p}$ decreases. However, even for the high value of the pseudopotential ($\mu^{\star}=0.3$) occurs $\omega_{p}>\Omega_{{\rm max}}$. The above result is related to the fact that the electron-phonon coupling constant for ${\rm Si_{2}H_{6}}$ takes the high value.
 
Basing on the presented results, the low-temperature value of the energy gap $2\Delta\left(0\right)$ at the Fermi level ($T=T_{0}$) has been calculated. In particular, the following expression has been used: 
$\Delta\left(T\right)={\rm Re}\left[\Delta\left(\omega=\Delta\left(T\right)\right)\right]$. As a result it has been obtained: 
$2\Delta\left(0\right)\in\left<65.73,34.85\right>$ meV for $\mu^{\star}\in\left<0.1,0.3\right>$. 

The knowledge of the energy gap's value allows to determine the dimensionless ratio $R_{\Delta}\equiv 2\Delta\left(0\right)/k_{B}T_{C}$. For compound ${\rm Si_{2}H_{6}}$ it has been achieved: $R_{\Delta}\in\left<4.40,4.04\right>$. The above result is really far from the $R_{\Delta}$ predicted by the BCS theory; $\left[R_{\Delta}\right]_{\rm BCS}=3.53$ \cite{BCS1}, \cite{BCS2}. Let us notice that the dependence of $R_{\Delta}$ on $\mu^{\star}$ can be determined with a good accuracy based on an analytical formula derived for the compound ${\rm SiH_4(H_2)_2}$ \cite{wzor2}. 
  
%
%

%
%

Second solution of the Eliashberg equations allows to determine the electron effective mass: 
$m^{\star}_{e}={\rm Re}\left[Z\left(\omega=0\right)\right] m_{e}$, where symbol $m_{e}$ stands for the band mass. After the appropriate calculations, it has been stated that the quantity $m^{\star}_{e}$ weakly depends on the temperature and the Coulomb pseudopotential. The electron effective mass reaches its maximum for $T=T_{C}$, and the value $2.397m_{e}$ has been obtained.

\vspace*{0.5cm}

The basic thermodynamic parameters characterizing the superconducting state in ${\rm Si_{2}H_{6}}$ under the pressure at $275$ GPa have been determined. 

It has been found that the critical temperature is very high even for the large values of the Coulomb pseudopotential: 
$T_{C}\in\left<173.36, 99.99\right>$ K.

In next step, the low-temperature value of the energy gap at the Fermi level has been determined. On the basis of the presented results, it has been proven that the dimensionless parameter $R_{\Delta}$ greatly exceeds the value predicted by the BCS theory: $R_{\Delta}\in\left<4.40, 4.04\right>$. 

In the last step, the electron effective mass has been calculated. It has been shown that in the whole range of the superconducting state's existence, the value of $m^{\star}_{e}$ is high and reaches the maximum in the critical temperature: $\left[m^{\star}_{e}\right]^{\rm max}_{T=T_{C}}=2.397m_{e}$.

\begin{acknowledgments}
The authors wish to thank Prof. K. Dzili{\'n}ski for providing excellent working conditions and the financial support.\\
Some calculations have been conducted on the Cz{\c{e}}stochowa University of Technology cluster, built in the framework of the
PLATON project, no. POIG.02.03.00-00-028/08 - the service of the campus calculations U3.  
\end{acknowledgments}
%

%

\begin{thebibliography}{99}
%
\bibitem{Nagamatsu}
J. Nagamatsu, N. Nakagawa, T. Muranaka, Y. Zenitani, A. Akimitsu, Nature {\bf 410}, 63 (2001).
\bibitem{Golubov}
A.A. Golubov, J. Kortus, O.V. Doglov, O. Jepsen, Y. Kong, O.K. Andersen, B.J. Gibson, K. Ahn, R.K. Kremer, J. Phys.: Condens.
Matter {\bf 14}, 1353 (2002).
\bibitem{Szczesniak1}
R. Szcz{\c{e}}{\'s}niak, Solid State Commun. {\bf 145}, 137 (2008).
\bibitem{BCS1}
J. Bardeen, L.N. Cooper, J.R. Schrieffer, Phys. Rev. {\bf 106}, 162 (1957).
\bibitem{BCS2}
J. Bardeen, L.N. Cooper, J.R. Schrieffer, Phys. Rev. {\bf 108}, 1175 (1957).
\bibitem{Bednorz}
J.G. Bednorz, K.A. Muller, Z. Phys. B {\bf 64}, 189 (1986).
\bibitem{Chu}
C.W. Chu, L. Gao, F. Chen, Z.J. Huang, R.L. Meng, Y.Y. Xue, Nature {\bf 365}, 323 (1993).
\bibitem{GaoHTSC}
L. Gao, Y.Y. Xue, F. Chen, Q. Xiong, R.L. Meng, D. Ramirez, C.W. Chu, J.H. Eggert, H.K. Mao, Phys. Rev. B {\bf 50}, 4260 (1994).
\bibitem{Szczesniak2}
R. Szcz{\c{e}}{\'s}niak, PLoS ONE 7 (4), art. no. e31873 (2012);  
preprint: arXiv:1105.5525 (2011) and arXiv:1110.3404 (2012).
\bibitem{Deemyad}
S. Deemyad, J.S. Schilling, Phys. Rev. Lett. {\bf 91}, 167001 (2003).
\bibitem{Yabuuchi}
T. Yabuuchi, T. Matsuoka, Y. Nakamoto, K. Shimizu, J. Phys. Soc. Jpn. {\bf 75}, 083703 (2006).
\bibitem{Sakata}
M. Sakata, Y. Nakamoto, K. Shimizu, T. Matsuoka, Y. Ohishi, Phys. Rev. B {\bf 83}, 220512(R) (2011).
\bibitem{Andersson}
M. Andersson, Phys. Rev. B {\bf 84}, 216501 (2011).
\bibitem{Szczesniak3}
R. Szcz{\c{e}}{\'s}niak, M.W. Jarosik, D. Szcz{\c{e}}{\'s}niak, Physica B {\bf 405}, 4897 (2010).
\bibitem{Szczesniak4}
R. Szcz{\c{e}}{\'s}niak, A.P. Durajski, Physica C {\bf 472}, 15 (2012).
\bibitem{Szczesniak5}
R. Szcz{\c{e}}{\'s}niak, A.P. Durajski, Journal of Superconductivity and Novel Magnetism {\bf 25}, 399 (2012).
\bibitem{Szczesniak6}
R. Szcz{\c{e}}{\'s}niak, A.P. Durajski, M.W. Jarosik, Mod. Phys. Lett. B {\bf 26}, 1250050 (2012). 
\bibitem{Szczesniak7}
R. Szcz{\c{e}}{\'s}niak, A.P. Durajski, Solid State Commun. {\bf 152}, 1018 (2012).
\bibitem{Szczesniak8}
R. Szcz{\c{e}}{\'s}niak, D. Szcz{\c{e}}{\'s}niak, Solid State Commun. {\bf 152}, 779 (2012). 
\bibitem{Ashcroft}
N.W. Ashcroft, Phys. Rev. Lett. {\bf 21}, 1748 (1968).
\bibitem{Cudazzo}
P. Cudazzo, G. Profeta, A. Sanna, A. Floris, A. Continenza,
S. Massidda, E.K.U. Gross, Phys. Rev. Lett. {\bf 100}, 257001 (2008).
\bibitem{Szczesniak9}
R. Szcz{\c{e}}{\'s}niak, M.W. Jarosik, Physica B {\bf 406}, 3493 (2011).
\bibitem{Szczesniak10}
R. Szcz{\c{e}}{\'s}niak, M.W. Jarosik, Physica B {\bf 406}, 2235 (2011).
\bibitem{Maksimov}
E.G. Maksimov, D.Y. Savrasov, Solid State Commun. {\bf 119}, 569 (2001).
\bibitem{McMahon}
J.M. McMahon, D.M. Ceperley, Phys. Rev. B {\bf 84}, 144515 (2011).
\bibitem{Szczesniak11}
R. Szcz{\c{e}}{\'s}niak, M.W. Jarosik, Solid State Commun. {\bf 149}, 2053 (2009).
\bibitem{Stadele}
M. Stadele, R.M. Martin, Phys. Rev. Lett. {\bf 84}, 6070 (2000).
\bibitem{Feng}
J. Feng, W. Grochala, T. Jaron, R. Hoffmann, A. Bergara, N.W. Ashcroft, Phys. Rev. Lett. {\bf 96}, 017006 (2006).
\bibitem{Chen}
X.J. Chen, V.V. Struzhkin, Y. Song, A.F. Goncharov, M. Ahart, Z. Liu, H. Mao, R.J. Hemley, 
Proc. Nat. Acad. Sci. USA {\bf 105}, 20 (2008).
\bibitem{Eremets}
M.I. Eremets, I.A. Trojan, S.A. Medvedev, J.S. Tse, Y. Yao, Science {\bf 319}, 1506 (2008).
\bibitem{Jin}
X. Jin, X. Meng, Z. He, Y. Ma, B. Liu, T. Cui, G. Zou, H. Mao, Proc. Nat. Acad. Sci. USA {\bf 107}, 9969 (2010).
\bibitem{Li}
Y. Li, G. Gao, Y. Xie, Y. Ma, T. Cui, G. Zou, Proc. Nat. Acad. Sci. USA {\bf 107}, 15708 (2010).
\bibitem{Eliashberg}
G.M. Eliashberg, Soviet. Phys. JETP {\bf 11}, 696 (1960).
\bibitem{Allen}
P.B. Allen, B. Mitrovi{\'c}, in: Solid State Physics: Advances in Research and Applications, edited by H. Ehrenreich, F. Seitz, D. Turnbull, 
(Academic, New York, 1982), Vol 37, p. 1.
\bibitem{Carbotte01}
J.P. Carbotte, Rev. Mod. Phys. {\bf 62}, 1027 (1990).
\bibitem{Carbotte02}
J.P. Carbotte, F. Marsiglio, in: The Physics of Superconductors, edited by K.H. Bennemann, J.B. Ketterson, (Springer, Berlin, 2003), Vol 1, p. 223.
\bibitem{Marsiglio}
F. Marsiglio, M. Schossmann, J.P. Carbotte, Phys. Rev. B {\bf 37}, 4965 (1988).
\bibitem{Szczesniak12}
R. Szcz{\c{e}}{\'s}niak, Solid State Commun. {\bf 144}, 27 (2007).
\bibitem{Szczesniak13}
R. Szcz{\c{e}}{\'s}niak, Phys. Lett. A {\bf 373}, 473 (2009).
\bibitem{Szczesniak14}
M.W. Jarosik, R. Szcz{\c{e}}{\'s}niak, D. Szcz{\c{e}}{\'s}niak, Acta Phys. Pol. A {\bf 118}, 1031 (2010).
\bibitem{Szczesniak15}
A.P. Durajski, R. Szcz{\c{e}}{\'s}niak, M.W. Jarosik, Phase Transitions, DOI:10.1080/01411594.2012.658051
\bibitem{AllenDynes}
P.B. Allen, R.C. Dynes, Phys. Rev. B {\bf 12}, 905 (1975).
\bibitem{McMillan}
W.L. McMillan, Phys. Rev. {\bf 167}, 331 (1968). 
\bibitem{wzor1}
In unpublished yet paper for ${\rm SiH_4(H_2)_2}$ (R. Szcz{\c{e}}{\'s}niak and A.P. Durajski), the formula has been given: 
$k_{B}T_{C}=f_{1}f_{2}\frac{\omega_{{\rm ln}}}{1.37}\exp\left[\frac{-1.125\left(1+\lambda\right)}{\lambda-\mu^{\star}}\right]$,
where $f_{1}$ and $f_{2}$ denote the correction functions \cite{AllenDynes}. The quantities 
$\Lambda_{1}$ and $\Lambda_{2}$ in $f_{1}$ i $f_{2}$ have the form:
$\Lambda_{1}=2-0.14\mu^{\star}$ and $\Lambda_{2}=\left(0.27+10\mu^{\star}\right)\left(\sqrt{\omega_{2}}/\omega_{\ln}\right)$. 
\bibitem{Varelogiannis}
G. Varelogiannis, Z. Phys. B {\bf 104}, 411 (1997). 
\bibitem{spirale}
X.H. Zheng, D.G. Walmsley, Phys. Rev. B {\bf 76}, 224520 (2007).
\bibitem{wzor2}
In unpublished yet paper for ${\rm SiH_4(H_2)_2}$ (R. Szcz{\c{e}}{\'s}niak and A.P. Durajski), the formula has been given: 
$\frac{R_{\Delta}}{\left[R_{\Delta}\right]_{{\rm BCS}}}=1+\left(\frac{k_{B}T_{C}}{a\omega_{{\rm ln}}}\right)^{2}
\left[\ln\left(\frac{a\omega_{{\rm \ln}}}{k_{B}T_{C}}\right)+\ln^{2}\left(\frac{a\omega_{{\rm \ln}}}{k_{B}T_{C}}\right)
\right]$, where $a=0.3447$.
\end{thebibliography}
\end{document}